# Soliton solutions and traveling wave solutions for the two-dimensional generalized nonlinear Schrödinger equations

Cestmir Burdik[1], Gaukhar Shaikhova[2], Berik Rakhimzhanov[3]

[1]Faculty of Nuclear Sciences and Physical Engineering,
CTU, Prague, Czech Republic
[2]Department of General & Theoretical Physics, Eurasian National University,
Nur-Sultan, Kazakhstan
[3]"JSC National company "Kazakhstan Gharysh Sapary", Nur-Sultan, Kazakhstan

E-mails: cestmir.burdik@fjfi.cvut.cz, g.shaikhova@gmail.com, rahimzhanovberik@gmail.com

**Abstract**
In this paper, we present the two-dimensional generalized nonlinear Schrödinger equations with the Lax pair. These equations are related to many physical phenomena in the Bose-Einstein condensates, surface waves in deep water and nonlinear optics. The existence of the Lax pair defines integrability for the partial differential equation, so the two-dimensional generalized nonlinear Schrödinger equations are integrable. We obtain bilinear forms of the two-dimensional GNLS equations. One- and two- soliton solutions are derived via the Hirota bilinear method, a procedure quite useful in the solution of nonlinear partial differential equations. We apply the extended tanh method in order to construct new exact traveling wave solutions. Through 3D plots, we show the dynamical behavior of the obtained solutions.

**Keywords:** soliton, traveling wave, generalized nonlinear Schrödinger equation, Hirota method, extended tanh method

## 1. Introduction

The investigation of nonlinear evolution equations is the main area of research in the field of nonlinear dynamics. One of the nonlinear equations is the nonlinear Schrödinger (NLS) equation which arises from a wide variety of fields, such as weakly nonlinear dispersive water waves, quantum field theory and nonlinear optics [1]–[4]. Different modifications and generations of the NLS equations were proposed and studied [5]–[10]. There are various methods to study nonlinear equations, such as the Darboux transformation [11]-[15], the Hirota method [16]-[19], the sine-cosine [20], [21], the extended tanh method [23]-[25] and so on.

In this paper, we present the two-dimensional generalized nonlinear Schrödinger equations (GNLSE) as

$$iq_t + q_{xy} - vq + \alpha q - i\beta q_x = 0, \qquad (1)$$
$$v_x + 2(|q|^2)_y = 0, \qquad (2)$$

where $q, v$ are the wave functions, $\alpha$ and $\beta$ are the constants. The equations (1)-(2) admit next reductions: if $\alpha = 0, \beta = 0$ we can obtain the two-dimensional nonlinear Schrödinger equations, if $x = y, \alpha = 0, \beta = 0$ we can get the one-dimensional nonlinear Schrödinger equation.

By using two methods we obtain nonlinear wave solutions for the two-dimensional GNLSE. We apply Hirota bilinear method and obtain the bilinear form of the two-dimensional GNLSE. One soliton and two soliton solutions are constructed based on the obtained bilinear form. We derive a traveling wave solution using the extended tanh-method that provides wider applicability for handling nonlinear wave equations. The figures have been plotted to display the dynamical features of solutions.

The article is organized as follows. In Sec. II, we present the Lax pair for the two-dimensional GNLSE. The Hirota bilinear method was applied for two-dimensional GNLSE (1)-(2) to obtain exact soliton solutions in Sec. III. Section IV is the derivation of a traveling wave solution for the two-dimensional GNLSE (1)-(2) by the extended tanh method. In Sec. V, we summarize the results of our study.

## 2. Lax pair

The Lax pair provides the complete integrability of the nonlinear equation. In this section, we present the Lax pair for equations (1)-(2) that can be expressed as follows:

$$\Psi_x = U\Psi, \tag{3}$$
$$\Psi_t = 2\lambda\Psi_y + V\Psi, \tag{4}$$

where $\Psi = (\Psi_1, \Psi_2)^T$ ($T$ denotes the transpose of a matrix), $\lambda$ is a spectral parameter and the matrices $U$ and $V$ have the form

$$U = \begin{pmatrix} -i\lambda & q \\ -r & i\lambda \end{pmatrix}, \tag{5}$$

$$V = \begin{pmatrix} -i\lambda\beta - \frac{iv}{2} + \frac{i\alpha}{2} & iq_y + \beta q \\ ir_y - \beta r & i\lambda\beta + \frac{iv}{2} - \frac{i\alpha}{2} \end{pmatrix}. \tag{6}$$

Through direct computations, it can be verified that the compatibility condition (also known as a zero- curvature condition):

$$U_t - V_x - 2\lambda U_y + UV - VU = 0, \tag{7}$$

exactly gives rise to

$$iq_t + q_{xy} - vq + \alpha q - i\beta q_x = 0,$$
$$ir_t - r_{xy} + vr - \alpha r - i\beta r_x = 0,$$
$$v_x + 2(rq)_y = 0.$$

Let us now consider the reduction $r = q^*$, where $*$ means a complex conjugate we use to get equations (1)-(2).

## 3. Soliton solutions

In order to obtain soliton solutions for the two-dimensional GNLSE we apply Hirota bilinear method. The method was suggested by Hirota [16],[18]. This approach provides a direct method for finding N-soliton solutions to nonlinear evolutionary equations. The stages of the method are described in the next section.

### 3.1 Description of Hirota bilinear method

The basic idea in Hirota bilinear method are as follows [3],[16],[18]:

 *Bilinearization.* At this stage, a dependent variable transformation is introduced. The transformation ought to reduce the nonlinear equation to the bilinear equation, which is quadratic in the dependent variables.

 *Transformation to the Hirota bilinear form.* Hirota suggests the D-operator defined by

$$D_x^l D_y^m D_t^n (g \cdot f) = \left(\frac{\partial}{\partial x} - \frac{\partial}{\partial x'}\right)^l \left(\frac{\partial}{\partial y} - \frac{\partial}{\partial y'}\right)^m \left(\frac{\partial}{\partial t} - \frac{\partial}{\partial t'}\right)^n g(x,y,t) \cdot f(x',y',t')|_{x=x', y=y', t=t'} \quad (8)$$

with $x', y'$ and $t'$ as three formal variables, $g(x,y,t)$ and $f(x',y',t')$ being two functions, $l, m$ and $n$ being three nonnegative integers. The operator (8) rewrites the bilinear equation in terms of the $D$ operator as a combination of variable coefficient bilinear equations.

 *Using the Hirota perturbation.* Formal perturbation expansion into this bilinear equation is introduced. This expansion is truncated in the case of soliton solutions. To prove that the suggested soliton form is indeed correct, we use mathematical induction.

### 3.2 Application

*Bilinear form*

The two-dimensional GNLSE (1)-(2) can be rewritten as

$$[iD_t + D_x D_y + \alpha - i\beta D_x](g \cdot f) = 0, \quad (9)$$
$$D_x D_y (f \cdot f) + 2D_y (h \cdot f) = 0, \quad (10)$$
$$D_x (h \cdot f) + |g|^2 = 0, \quad (11)$$

with the dependent variable transformations

$$q = \frac{g}{f}, \quad (12)$$
$$v = 2\left(\frac{h}{f}\right)_y, \quad (13)$$

where $g$ is the complex function of $x, y$ and $t$; $f, h$-are real ones, $D_x, D_y$ and $D_t$ are the bilinear differential operators defined by (8).

To obtain the soliton solutions of equations (9)-(11), we expand $g, f, h$ with respect to a small parameter $\varepsilon$ as follows:

$$g(x,y,t) = \varepsilon g_1(x,y,t) + \varepsilon^3 g_3(x,y,t) + \cdots, \quad (14)$$
$$f(x,y,t) = 1 + \varepsilon^2 f_2(x,y,t) + \varepsilon^4 f_4(x,y,t) + \cdots, \quad (15)$$
$$h(x,y,t) = 1 + \varepsilon^2 h_2(x,y,t) + \varepsilon^4 h_4(x,y,t) + \cdots, \quad (16)$$

where $g_j$ ($j = 1,3,5, \ldots$) are the complex functions of $x, y$ and $t$, and $f_n, h_n$ ($n = 2,4,6, \ldots$) are the real ones. Substituting expression (14)-(16) into (9)-(11) and collecting the coefficients of the same power of $\varepsilon$, we have from equation (9)

$$\varepsilon^1: [iD_t + D_xD_y + \alpha - i\beta D_x](g_1 \cdot 1) = 0,$$
$$\varepsilon^3: [iD_t + D_xD_y + \alpha - i\beta D_x](g_3 \cdot 1 + g_1 \cdot f_2) = 0,$$
$$\varepsilon^5: [iD_t + D_xD_y + \alpha - i\beta D_x](g_5 \cdot 1 + g_3 \cdot f_2 + g_1 \cdot f_4) = 0,$$
$$\varepsilon^7: [iD_t + D_xD_y + \alpha - i\beta D_x](g_5 \cdot f_2 + g_3 \cdot f_4) = 0$$
...,

from equation (10)

$$\varepsilon^2: D_xD_y(f_2 \cdot 1 + 1 \cdot f_2) + 2D_y(h_2 \cdot 1 + 1 \cdot f_2) = 0,$$
$$\varepsilon^4: D_xD_y(f_4 \cdot 1 + f_2 \cdot f_2 + 1 \cdot f_4) + 2D_y(h_4 \cdot 1 + h_2 \cdot f_2 + 1 \cdot f_4) = 0,$$
$$\varepsilon^6: D_xD_y(f_4 \cdot f_2 + f_2 \cdot f_4) + 2D_y(h_4 \cdot f_2 + h_2 \cdot f_4) = 0,$$
...

and from equation (11)
$$\varepsilon^2: D_x(h_2 \cdot 1 + 1 \cdot f_2) + g_1^* g_1 = 0,$$
$$\varepsilon^4: D_x(h_4 \cdot 1 + h_2 \cdot f_2 + 1 \cdot f_4) + g_3 g_1^* + g_1 g_3^* = 0,$$
$$\varepsilon^6: D_x(h_4 \cdot f_2 + h_2 \cdot f_4) + g_5 g_1^* + g_3 g_3^* + g_5 g_1^* = 0,$$
...

with the benefit of the above expression and symbolic computation, we can obtain the one-, two-, and N-soliton solutions for equations (1)-(2).

*The one-soliton solutions*
Truncating expressions (14)-(16) as

$$g(x,y,t) = \varepsilon g_1(x,y,t), f(x,y,t) = 1 + \varepsilon^2 f_2(x,y,t), h(x,y,t) = 1 + \varepsilon^2 h_2(x,y,t), \quad (17)$$

setting $\varepsilon = 1$, and substituting them into bilinear forms (9)-(11), we can obtain the one-soliton solutions for the two-dimensional GNLSE as follows:

$$q = \frac{e^{\theta_1}}{1+e^{\theta_1+\theta_1^*+R}}, \quad (18)$$
$$v = 2(\frac{1+e^{\theta_1+\theta_1^*+S}}{1+e^{\theta_1+\theta_1^*+R}})_y, \quad (19)$$

where
$e^R = \frac{1}{(k_1+k_1^*)^2}, \quad e^S = \frac{1-k_1-k_1^*}{(k_1+k_1^*)^2}, \quad \theta_1 = k_1 x + p_1 y + (-\beta k_1 + ik_1 p_1 + i\alpha)t + \theta_{10}$

with $k_1, p_1, \theta_{10}$ as constants. Figures 1 and 2 show the time evolutions of the one-soliton solutions.

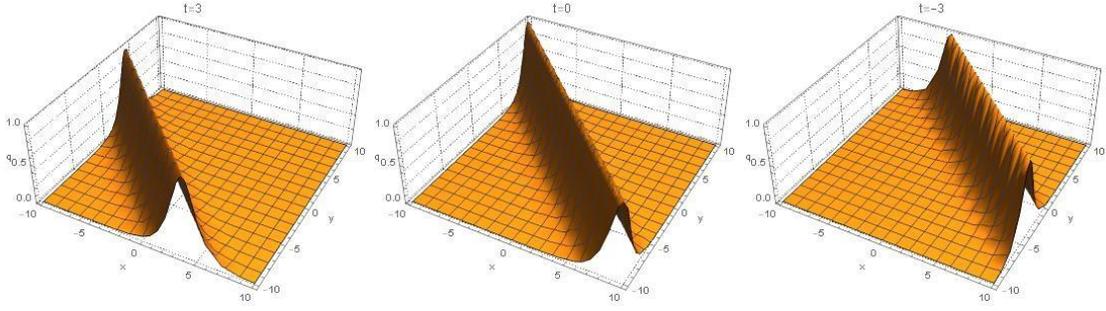

Figure 1: The time evolutions of the one-soliton solution (18).
The parameters are $k_1 = 1 + I; p_1 = 1 - I; \alpha = 1; \beta = 2$

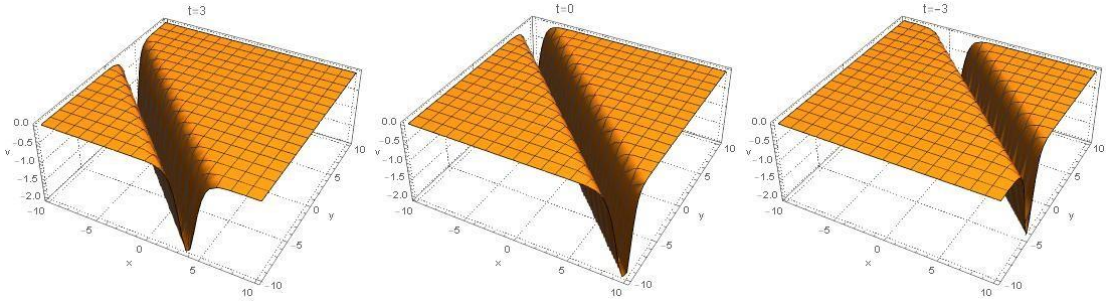

Figure 2: The time evolutions of the one-soliton solution (19).
The parameters are $k_1 = 1 + I; p_1 = 1 - I; \alpha = 1; \beta = 2$

*The two-soliton solutions*

To derive the two-soliton solutions for equations (1)-(2), we truncate expressions (14)-(16) as

$$g(x,y,t) = \varepsilon g_1(x,y,t) + \varepsilon^3 g_3(x,y,t), \tag{20}$$
$$f(x,y,t) = 1 + \varepsilon^2 f_2(x,y,t) + \varepsilon^4 f_4(x,y,t), \tag{21}$$
$$h(x,y,t) = 1 + \varepsilon^2 h_2(x,y,t) + \varepsilon^4 h_4(x,y,t), \tag{22}$$

set $\varepsilon = 1$, and substitute them into the bilinear equation (9)-(11), and we get

$$q = \frac{g_1 + g_3}{1 + f_2 + f_4}, \tag{23}$$

$$v = 2\left(\frac{1 + h_2 + h_4}{1 + f_2 + f_4}\right)_y, \tag{24}$$

where

$$g_1 = e^{\theta_1} + e^{\theta_2},$$

$$g_3 = e^{\theta_1+\theta_2+\theta_2^*+\delta_1} + e^{\theta_1+\theta_2+\theta_1^*+\delta_2},$$

$$f_2 = e^{\theta_1+\theta_1^*+R_{11}} + e^{\theta_2+\theta_1^*+R_{21}} + e^{\theta_1+\theta_2^*+R_{21}^*} + e^{\theta_2+\theta_2^*+R_{22}},$$

$$f_4 = e^{\theta_1+\theta_1^*+\theta_2+\theta_2^*+R_3},$$

$$h_2 = e^{\theta_1+\theta_1^*+S_{11}} + e^{\theta_2+\theta_1^*+S_{21}} + e^{\theta_1+\theta_2^*+S_{21}^*} + e^{\theta_2+\theta_2^*+S_{22}},$$

$$h_4 = e^{\theta_1+\theta_1^*+\theta_2+\theta_2^*+S_3}$$

with

$$e^{R_{11}} = \frac{1}{(k_1+k_1^*)^2}, \quad e^{R_{21}} = \frac{1}{(k_2+k_1^*)^2}, \quad e^{R_{22}} = \frac{1}{(k_2+k_2^*)^2},$$

$$e^{S_{11}} = \frac{1-k_1-k_1^*}{(k_1+k_1^*)^2}, \quad e^{S_{21}} = \frac{1-k_2-k_1^*}{(k_2+k_1^*)^2}, \quad e^{S_{22}} = \frac{1-k_2-k_2^*}{(k_2+k_2^*)^2},$$

$$e^{\delta_1} = -\frac{e^{R_{21}}((k_2-k_1)(p_1^*+p_2)+(k_2+k_1^*)(p_2-p_1))}{((k_2+k_1^*)(p_1^*+p_1)+(k_1+k_1^*)(p_1^*+p_2))} + \frac{e^{R_{11}}((k_1+k_1^*)(p_2-p_1)+(k_2-k_1)(p_1+p_1^*))}{((k_2+k_1^*)(p_1^*+p_1)+(k_1+k_1^*)(p_1^*+p_2))},$$

$$e^{\delta_2} = -\frac{e^{R_{22}}((k_2-k_1)(p_2^*+p_2)+(k_2+k_2^*)(p_2-p_1))}{((k_2+k_2^*)(p_2^*+p_1)+(k_1+k_2^*)(p_2^*+p_2))} + \frac{e^{R_{21}^*}((k_1+k_2^*)(p_2-p_1)+(k_2-k_1)(p_1+p_2^*))}{((k_2+k_2^*)(p_2^*+p_1)+(k_1+k_2^*)(p_2^*+p_2))},$$

$$e^{R_3} = \frac{(e^{\delta_1}+e^{\delta_2}+e^{\delta_1^*}+e^{\delta_2^*})}{(k_1+k_2+k_1^*+k_2^*)^2},$$

$$e^{R_3} = \frac{(1-k_1-k_2-k_1^*-k_2^*)(e^{\delta_1}+e^{\delta_2}+e^{\delta_1^*}+e^{\delta_2^*})}{(k_1+k_2+k_1^*+k_2^*)^2},$$

$$\theta_1 = k_1 x + p_1 y + (-\beta k_1 + ik_1 p_1 + i\alpha)t + \theta_{10},$$
$$\theta_2 = k_2 x + p_2 y + (-\beta k_2 + ik_2 p_2 + i\alpha)t + \theta_{20}.$$

Dynamics of the two-soliton solutions is presented (Fig.3, Fig. 4).

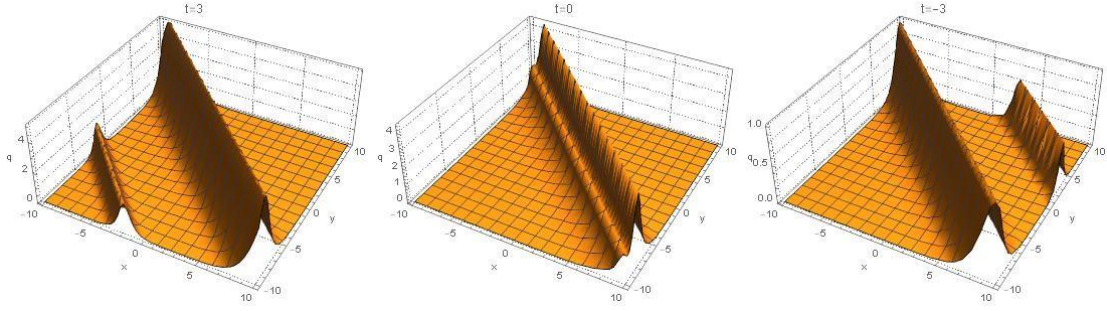

Figure 3: The time evolutions of the two-soliton solution (23).
The parameters are $k_1 = 1 + I; p_1 = 1 + I; k_2 = 2 - I; p_2 = 2 - I; α = 1; β = 2$

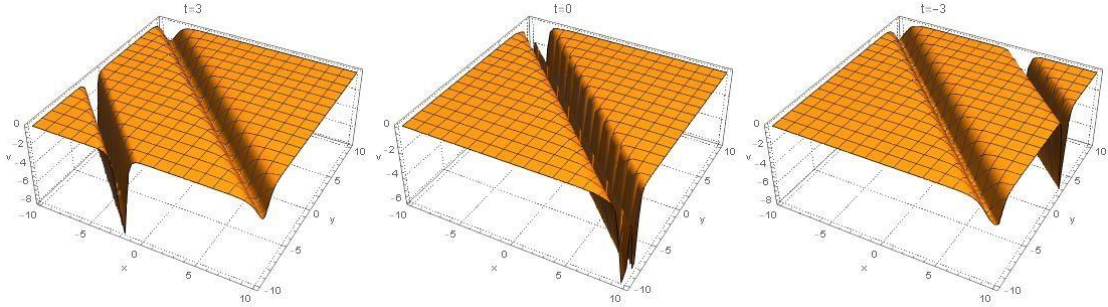

Figure 4: The time evolutions of the two-soliton solution (24).
The parameters are $k_1 = 1 + I; p_1 = 1 + I; k_2 = 2 - I; p_2 = 2 - I; α = 1; β = 2$

## 4. Traveling wave solutions

We use the extended tanh method [23] to obtain traveling wave solutions for the two-dimensional GNLSE. In the next section, the description of the method is presented.

### *4.1 Description of the extended tanh method*

The partial differential equation (PDE)

$$E_1(q, q_t, q_x, q_y, q_{tt}, q_{xx}, q_{yy}, \dots) = 0, \tag{25}$$

where $E_1$ is a polynomial of $q(x, y, t)$ and its partial derivatives, in which the highest order derivatives and nonlinear terms are involved, can be converted to the ordinary differential equation (ODE)

$$E_2(Q, Q', Q'', Q''', \dots) = 0, \tag{26}$$

by using a wave variable
$$q(x,y,t) = Q(\xi), \xi = x + y - ct, \tag{27}$$

where $c$ is the constant. We integrate equation (26) as long as all terms contain derivatives. Constants of integration are considered zeros. The tanh method was suggested by Maliet [22] and then was extended by Wazwaz [23]. By using a new independent variable
$$Y = tanh(\mu\xi), \quad \xi = x + y - ct, \tag{28}$$

where $\mu$ is the wave number, we have the following change of derivatives:
$$\frac{d}{d\xi} = \mu(1 - Y^2)\frac{d}{dY},$$
$$\frac{d^2}{d\xi^2} = -2\mu^2 Y(1 - Y^2)\frac{d}{dY} + \mu^2(1 - Y^2)\frac{d^2}{dY^2}.$$

The extended tanh method admits the use of the finite expansion in the following form:
$$Q(\xi) = \sum_{n=0}^{M} a_n Y^n + \sum_{n=1}^{M} b_n Y^{-n}, \tag{29}$$

where $a_0, a_1, a_2, a_3 \ldots a_N$ and $b_0, b_1, b_2, b_3 \ldots b_N$ are unknown constants. $M$ is obtained balancing the highest order derivative term and the non-linear terms in equation (26). Then put the value of $Q(\xi)$ from (29) in equation (26), and comparing the coefficient of $Y^n$ we can obtain the values of the coefficients where $a_0, a_1, a_2, a_3 \ldots a_N$ and $b_0, b_1, b_2, b_3 \ldots b_N$.

## 4.2 Application

In this section, we obtain exact traveling wave solutions of the two-dimensional GNLS equation using the extended tanh method [23],[25]. For applying this method, we ought to reduce the system (1)-(2) to the system of ordinary deferential equations. If we consider the transformation
$$q(x,y,t) = e^{i(ax+by+dt)} Q(x,y,t), \tag{30}$$

where $a, b, d$ are the constants, $Q(x, y, t)$ is the real valued function, then the system (1)-(2) reduced to the following system of differential equations
$$Q(-d - ba + \alpha + \beta a) + Q_{xy} - vQ = 0, \tag{31}$$
$$Q_t + aQ_y + (b - \beta)Q_x = 0, \tag{32}$$
$$v_x + 2(Q^2)_y = 0. \tag{33}$$

Substituting the wave transformation
$$Q(x,y,t) = Q(\xi) = Q(x + y - ct), \tag{34}$$
$$v(x,y,t) = V(\xi) = V(x + y - ct), \tag{35}$$

into system (31)-(33), we obtain that

$$Q(-d - ba + \alpha + \beta a) + Q'' - VQ = 0, \tag{36}$$
$$Q'(-c + a + b - \beta) = 0, \tag{37}$$
$$V' + 2(Q^2)' = 0. \tag{38}$$

From equation (37) we have that

$$c = a + b - \beta. \tag{39}$$

Integrating equation (38) with respect to $\xi$ and taking integration constant zero for simplicity, we find
$$V = -2Q^2. \tag{40}$$
Substituting equation (40) into equation (36), we obtain the following ordinary differential equation

$$Q(-d + (\beta - b)a + \alpha) + Q'' + 2Q^3 = 0, \tag{41}$$

where prime denotes the derivation with respect to $\xi$. Balancing the nonlinear term $Q^3$, which has the exponent $3M$, with the highest order derivative $Q''$, which has the exponent $M + 2$, in (41) yields $3M = M + 2$ that gives $M = 1$. Then the extended tanh method allows us to use the substitution

$$Q(\xi) = a_0 + a_1 Y + \frac{b_1}{Y}. \tag{42}$$

Substituting (42) into (41) and collecting the coefficients of $Y$, we obtain a system of algebraic equations for $a_0, a_1, b_1, \mu$. Solving this system with the aid of Maple, we obtain the following results:

*Result 1:*

$$a_0 = 0, c = a + b - \beta, \tag{43}$$
$$a_1 = \pm \frac{1}{2}\sqrt{\frac{1}{2}(a(b-\beta) - \alpha + d)}, \tag{44}$$
$$b_1 = \pm \frac{1}{2}\sqrt{\frac{1}{2}(a(b-\beta) - \alpha + d)}, \tag{45}$$
$$\mu_1 = \pm \frac{1}{2}\sqrt{-\frac{1}{2}(a(b-\beta) - \alpha + d)}. \tag{46}$$

*Result 2:*

$$a_0 = 0, \ a_1 = 0, c = a + b - \beta, \tag{47}$$
$$b_1 = \pm \sqrt{\frac{1}{2}(a(b-\beta) - \alpha + d)}, \tag{48}$$
$$\mu_1 = \pm \sqrt{-\frac{1}{2}(a(b-\beta) - \alpha + d)}. \tag{49}$$

*Result 3:*

$$a_0 = 0, \ b_1 = 0, c = a + b - \beta, \tag{50}$$

$$a_1 = \pm\sqrt{\tfrac{1}{2}(a(b-\beta) - \alpha + d)}, \tag{51}$$

$$\mu_1 = \pm\sqrt{-\tfrac{1}{2}(a(b-\beta) - \alpha + d)}. \tag{52}$$

*Result 4:*

$$a_0 = 0, \quad c = a + b - \beta, \tag{53}$$
$$a_1 = \mp\tfrac{1}{2}\sqrt{-ab + a\beta + \alpha - d}, \tag{54}$$
$$b_1 = \pm\tfrac{1}{2}\sqrt{-ab + a\beta + \alpha - d}, \tag{55}$$
$$\mu_1 = \pm\tfrac{1}{2}\sqrt{ab - a\beta - \alpha + d}. \tag{56}$$

By substituting equation (42) into (34),(40) and then the obtained expressions into (30) and (35), we can obtain new solutions for the two-dimensional GNLSE (1)-(2) in the following form

$$q(x,y,t) = e^{i(ax+by+dt)}[a_0 + a_1 \tanh(\mu\xi) + b_1 \coth(\mu\xi)], \tag{57}$$

$$v(x,y,t) = -2[a_0 + a_1 \tanh(\mu\xi) + b_1 \coth(\mu\xi)]^2, \tag{58}$$

where $\xi = x + y - ct$.
Finally, substituting the results (43)-(56) into (57)-(58), we can obtain traveling wave solutions in the next form

$$q_1(x,y,t) =$$
$$= e^{i(ax+by+dt)}\left[\pm\tfrac{1}{2}\sqrt{\tfrac{1}{2}(a(b-\beta)-\alpha+d)}\left(\tanh\left(\pm\tfrac{1}{2}\sqrt{-\tfrac{1}{2}(a(b-\beta)-\alpha+d)}\xi\right) + \coth\left(\pm\tfrac{1}{2}\sqrt{-\tfrac{1}{2}(a(b-\beta)-\alpha+d)}\xi\right)\right)\right],$$

$$v_1(x,y,t) = -2\left[\pm\tfrac{1}{2}\sqrt{\tfrac{1}{2}(a(b-\beta)-\alpha+d)}\left(\tanh\left(\pm\tfrac{1}{2}\sqrt{-\tfrac{1}{2}(a(b-\beta)-\alpha+d)}\xi\right) + \coth\left(\pm\tfrac{1}{2}\sqrt{-\tfrac{1}{2}(a(b-\beta)-\alpha+d)}\xi\right)\right)\right]^2,$$

$$q_2(x,y,t) = e^{i(ax+by+dt)}\left[\pm\sqrt{\frac{1}{2}(a(b-\beta)-\alpha+d)}\coth\left(\pm\sqrt{-\frac{1}{2}(a(b-\beta)-\alpha+d)}\xi\right)\right],$$

$$v_2(x,y,t) = -2\left[\pm\sqrt{\frac{1}{2}(a(b-\beta)-\alpha+d)}\coth\left(\pm\sqrt{-\frac{1}{2}(a(b-\beta)-\alpha+d)}\xi\right)\right]^2,$$

$$q_3(x,y,t) = e^{i(ax+by+dt)}\left[\pm\sqrt{\frac{1}{2}(a(b-\beta)-\alpha+d)}\tanh\left(\pm\sqrt{-\frac{1}{2}(a(b-\beta)-\alpha+d)}\xi\right)\right],$$

$$v_3(x,y,t) = -2\left[\pm\sqrt{\frac{1}{2}(a(b-\beta)-\alpha+d)}\tanh\left(\pm\sqrt{-\frac{1}{2}(a(b-\beta)-\alpha+d)}\xi\right)\right]^2,$$

$$q_4(x,y,t) = e^{i(ax+by+dt)}\left[\begin{array}{l}\mp\frac{1}{2}\sqrt{-ab+a\beta+\alpha-d}\tanh\left(\pm\frac{1}{2}\sqrt{ab-a\beta-\alpha+d}\,\xi\right)\pm\\ \pm\frac{1}{2}\sqrt{-ab+a\beta+\alpha-d}\coth\left(\pm\frac{1}{2}\sqrt{ab-a\beta-\alpha+d}\,\xi\right)\end{array}\right],$$

$$v_4(x,y,t) = -2\left[\begin{array}{l}\mp\frac{1}{2}\sqrt{-ab+a\beta+\alpha-d}\tanh\left(\pm\frac{1}{2}\sqrt{ab-a\beta-\alpha+d}\,\xi\right)\pm\\ \pm\frac{1}{2}\sqrt{-ab+a\beta+\alpha-d}\coth\left(\pm\frac{1}{2}\sqrt{ab-a\beta-\alpha+d}\,\xi\right)\end{array}\right]^2,$$

where $\xi = x + y - (a + b - \beta)t$.

## 5. Conclusion

In this paper, we present the two-dimensional generalized nonlinear Schrödinger equations with the Lax pair. The Lax pair plays an important role in the study of the integrability of the differential system. By employing two methods we have obtained the nonlinear wave solutions for the two-dimensional generalized nonlinear Schrödinger equations. Soliton solutions are derived by Hirota bilinear method. This method gives a mechanism for finding arbitrary N-soliton solutions for PDEs which can be written in bilinear form in the D-operator via a transformation of the dependent variable. We obtained the traveling wave solutions using the extended tanh method that provides wider applicability for handling nonlinear wave equations. The figures are plotted to display the dynamical features of those solutions. Moreover, the presented methods can be applied to obtain new solutions for other nonlinear evolution equations.

# Acknowledgment

G.Shaikhova was supported by the Ministry of Education and Science of the Republic of Kazakhstan (F.0811, No.0118RK00935)

# References

[1]. V. E. Zakharov, A B Shabat. Exact theory of two-dimensional self-focusing and one-dimensional selfmodulation of waves in nonlinear media. Sov. Phys.-JETP, 34(1):62-69, 1972. http://jetp.ac.ru/cgibin/dn/e-034-01-0062.

[2]. M J Ablowitz, P A Clarkson. Solitons, Nonlinear Evolution Equations and Inverse Scattering. First printing. Cambridge: Cambridge University Press, 1991.https://doi.org/10.1017/CBO9780511623998.

[3]. M.J. Ablowitz, H. Segur. Solitons and inverse scattering transform, Philadelphia: SIAM, 1981. DOI:0.1137/1.9781611970883.

[4]. M J Ablowitz , et al. Discrete and Continuous Nonlinear Schr•odinger Systems . Cambridge: Cambridge University Press, 2004. DOI: 10.2307/20453816.

[5]. K. Mio, et al. Modi_ed Nonlinear Schr•odinger Equation for Alfv_en Waves Propagating along the Magnetic Field in Cold Plasmas. J.Phys. Soc. Jpn. 41: 265-271, 1976. DOI:10.1143/JPSJ.41.265.

[6]. R. Guo, H. Hao. Breathers and multi-soliton solutions for the higher-order generalized nonlinear Schrodinger equation. Communications in Nonlinear Science and Numerical Simulation, 18(9): 2426-2435, 2013. DOI:10.1016/j.cnsns.2013.01.019.

[7]. S. Novikov, et al. Theory of solitons: the inverse scattering method. First printing. Springer Science and Business Media, 1984. https://www.springer.com/gp/book/9780306109775.

[8]. .E. Zakharov. Stability of periodic waves of _nite amplitude on the surface of a deep uid. J. Appl. Mech. Tech. Phys. 9(2):190-194, 1968. https://link.springer.com/article/10.1007/BF00913182.

[9]. M.S. Osman, et al. The uni_ed method for conformable time fractional Schr•odinger equation with perturbation terms. Chin. J.Phys. 56(5): 2500-2506, 2018. DOI:10.1016/j.cjph.2018.06.009.

[10]. J.J. Su, Y.T. Gao. N th-order bright and dark solitons for the higher-order nonlinear Schr•odinger equation in an optical _ber. Superlattices Microstruct. 120: 697-719, 2018. DOI: 10.1016/j.spmi.2017.12.020.

[11]. V.B. Matveev, M.A. Salle. Darboux Transformations and Solitons. First printing. Springer-Verlag Berlin Heidelberg, 1991. https://www.springer.com/gp/book/9783662009246.

[12]. G. Bekova, et al. Dark and bright solitons for the two-dimensional complex modi_ed Korteweg-de Vries and Maxwell-Bloch system with time-dependent coe_cient. Journal of Physics: Conference Series - 965: 012035, 2018. DOI: 10.1088/1742-6596/965/1/012035

[13]. Ch. Li, J. He Darboux transformation and positons of the inhomogeneous Hirota and the Maxwell-Bloch equation. Science China: Physics, Mechanics and Astronomy, 57(5): 898, 2014. DOI:10.1007/s11433-013-5296-x

[14]. K.R. Yesmakanova, et al. Exact solutions for the (2+1)-dimensional Hirota-Maxwell-Bloch system. AIP Conf. Proc. 1880:060022, 2017. DOI: 10.1063/1.5000676.

[15]. K.R. Yesmakanova, et al. Determinant Reprentation of Darboux Transformation for the (2+1)-Dimensional Schrodinger-Maxwell-Bloch Equation. Advances in Intelligent Systems and Computing, 441:183-198, 2016. DOI: 10.1007/978-3-319-30322-2-13.

[16]. R. Hirota Exact Solution of the Korteweg-de Vries Equation for Multiple Collisions of Solitons. Physical Review Letters, 27:1192-1194, 1971. DOI:/10.1103/PhysRevLett.27.1192.

[17]. K. Mukhanmedina, et al. Soliton solutions of two-component Hirota equation. Bulletin of the Karaganda University 80:103-107, 2015. https://mathematics-vestnik.ksu.kz/ru/srch/2015Mathematics-4-80-2015.

[18]. R. Hirota. The Direct Method in Soliton Theory. Cambridge University Press, Cambridge, 2005. DOI:10.1017/CBO9780511543043.

[19]. W. Hereman, A. Nuseir. Symbolic methods to construct exact solutions of nonlinear partial differential equations. Mathematics and Computers in Simulation 43: 13-27, 1997. DOI:10.1016/S0378-4754(96)00053-5.

[20]. M. T. Darvishi, et al. Travelling wave solutions for Boussinesq-like equations with spatial and spatial-temporal dispersion. Romanian Reports in Physics, 70(2):1-12, 2018. http://www.rrp.in_m.ro/IP/A333.

[21]. A.M. Wazwaz. The tanh and the sine-cosine methods for a reliable treatment of the modified equal width equation and its variants. Communications in Nonlinear Science and Numerical Simulation, 11(2):148-160, 2006. DOI:10.1016/j.cnsns.2004.07.001.

[22]. W. Malfiet, W. Hereman. The tanh method: I. Exact solutions of nonlinear evolution and wave equations. Physica Scripta, 54:563-568, 1996. DOI: 10.1088/0031-8949/54/6/003.

[23]. A.M. Wazwaz. Partial di_erential equations and solitary waves theory , Springer-Verlag Berlin Heidelberg, 2009. DOI:10.1007/978-3-642-00251-9.

[24]. G. Bekova , et al. Travelling wave solutions for the two-dimensional Hirota system of equations. AIP Conf. Proc. 1997: 020039, 2018. DOI: 10.1063/1.5049033.

[25]. A.M.Wazwaz. The extended tanh method for new solitons solutions for many forms of the fifth-order KdV equations. Appl Math Comput. 184(2):1002-1014, 2007. DOI:10.1016/j.amc.2006.07.002.